\documentclass{article} 
\usepackage{fullpage}
\usepackage{latexsym,amsmath,amssymb,color,rotating,xspace,epic,eepic,latexsym}
\usepackage{pstricks,pst-coil}
\usepackage{graphics,graphicx}
\newdimen \myunit
\newdimen \myhsize
\newdimen \myvsize
\newcommand{\wire}[1]{\raisebox{3\myunit}[0cm][0cm]{\small #1}}
\newcommand{\smwidth}{8}
\newcommand{\smhalf}{4}
\newcommand{\qthicklines}{\linethickness{1.5\myunit}}

\newcommand{\qheight}{10}
\newcommand{\qtopheight}{5}
\newcommand{\qtopheightminusthreehalves}{3.5}
\newcommand{\qtopheightplustwo}{7}
\newcommand{\qbottomheight}{5}
\newcommand{\qbottomheightminusthreehalves}{3.5}
\newcommand{\qbottomheightplustwo}{7}

\definecolor{darkgreen}{rgb}{0,0.5,0}

\def\cn#1{
\begin{picture}(4,\qheight)(0,0)
  \put(2,\qtopheight){\circle*{2}}
  \put(0,\qtopheight){\line(1,0){4}}
\ifcase #1
    \put(2,0){\line(0,1){\qheight}}
\or \put(2,\qtopheight){\line(0,1){\qbottomheight}}
\or \put(2,\qtopheight){\line(0,-1){\qtopheight}}
\fi
\end{picture}
}

\def\ccn#1#2{
\begin{picture}(4,\qheight)(0,0)
  \put(0,\qtopheight){\line(1,0){4}}
\textcolor{#1}{
  \put(2,\qtopheight){\circle*{2}}
\ifcase #2
    \put(2,0){\line(0,1){\qheight}}
\or \put(2,\qtopheight){\line(0,1){\qbottomheight}}
\or \put(2,\qtopheight){\line(0,-1){\qtopheight}}
\fi}
\end{picture}
}

\def\nn#1{
\begin{picture}(4,\qheight)(0,0)
  \put(2,\qtopheight){\circle{3}}
  \put(0,\qtopheight){\line(1,0){.5}}
  \put(4,\qtopheight){\line(-1,0){.5}}
\ifcase #1
    \put(2,0){\line(0,1){\qtopheightminusthreehalves}}
    \put(2,\qheight){\line(0,-1){\qbottomheightminusthreehalves}}
\or \put(2,\qheight){\line(0,-1){\qbottomheightminusthreehalves}}
\or \put(2,0){\line(0,1){\qtopheightminusthreehalves}}
\fi
\end{picture}
}

\def\cnn#1#2{
\begin{picture}(4,\qheight)(0,0)
  \textcolor{#1}{\put(2,\qtopheight){\circle{3}}}
  \put(0,\qtopheight){\line(1,0){.5}}
  \put(4,\qtopheight){\line(-1,0){.5}}
\textcolor{#1}{
\ifcase #2
    \put(2,0){\line(0,1){\qtopheightminusthreehalves}}
    \put(2,\qheight){\line(0,-1){\qbottomheightminusthreehalves}}
\or \put(2,\qheight){\line(0,-1){\qbottomheightminusthreehalves}}
\or \put(2,0){\line(0,1){\qtopheightminusthreehalves}}
\fi}
\end{picture}
}

\def\xn#1{
\begin{picture}(4,\qheight)(0,0)
  \put(2,\qtopheight){\circle{4}}
  \put(0,\qtopheight){\line(1,0){4}}
\ifcase #1
    \put(2,0){\line(0,1){\qheight}}
\or \put(2,\qheight){\line(0,-1){\qbottomheightplustwo}}
\or \put(2,0){\line(0,1){\qtopheightplustwo}}
\or \put(2,\qtopheightplustwo){\line(0,-1){4}}
\fi
\end{picture}
}

\def\cxn#1#2{
\begin{picture}(4,\qheight)(0,0)
  \textcolor{#1}{
  \put(2,\qtopheight){\circle{4}}
  \put(0,\qtopheight){\line(1,0){4}}
\ifcase #2
    \put(2,0){\line(0,1){\qheight}}
\or \put(2,\qheight){\line(0,-1){\qbottomheightplustwo}}
\or \put(2,0){\line(0,1){\qtopheightplustwo}}
\or \put(2,\qtopheightplustwo){\line(0,-1){4}}
\fi}
\end{picture}
}

\newcommand{\sn}[1]{
\begin{picture}(4,\qheight)(0,0)
\ifcase #1
    \put(0,\qtopheight){\line(1,0){4}}
\or \put(2,0){\line(0,1){\qheight}}
\or \put(0,\qtopheight){\line(1,0){4}}
    \put(2,0){\line(0,1){\qheight}}
\or \put(0,\qtopheight){\line(1,0){4}}
    \multiput(2,1)(0,\qtopheight){2}{\line(0,1){\qtopheightminusthreehalves}}
\fi
\end{picture}
}

\newcommand{\csn}[2]{
\begin{picture}(4,\qheight)(0,0)
\ifcase #2
    \put(0,\qtopheight){\line(1,0){4}}
\or \textcolor{#1}{\put(2,0){\line(0,1){\qheight}}}
\or \put(0,\qtopheight){\line(1,0){4}}
    \textcolor{#1}{\put(2,0){\line(0,1){\qheight}}}
\fi
\end{picture}
}

\def\sm#1{
\begin{picture}(\smwidth,\qheight)(0,0)
\ifcase #1
    \put(0,\qtopheight){\line(1,0){\smwidth}}
\or \put(\smhalf,0){\line(0,1){\qheight}}
\or \put(0,\qtopheight){\line(1,0){\smwidth}}
    \put(\smhalf,0){\line(0,1){\qheight}}
\or \put(0,\qtopheight){\line(1,0){\smwidth}}
    \multiput(\smhalf,1)(0,\qtopheight){2}{\line(0,1){\qtopheightminusthreehalves}}
\fi
\end{picture}
}

\def\sx#1{
\begin{picture}(12,\qheight)(0,0)
\ifcase #1
    \put(0,\qtopheight){\line(1,0){12}}
\or \put(6,0){\line(0,1){\qheight}}
\or \put(0,\qtopheight){\line(1,0){12}}
    \put(6,0){\line(0,1){\qheight}}
\or \multiput(0,\qtopheight)(2.6,0){\qtopheight}{\line(1,0){1.6}}
    \put(1,0){\line(0,1){\qheight}}
    \qthicklines
    \put(11,0){\line(0,1){\qheight}}
    \thinlines
\or \multiput(0,\qtopheight)(2.6,0){\qtopheight}{\line(1,0){1.6}}
    \qthicklines
    \put(1,0){\line(0,1){\qheight}}
    \thinlines
    \put(11,0){\line(0,1){\qheight}}
\fi
\end{picture}
}

\def\dx#1{
\begin{picture}(12,\qheight)(0,0)
  \put(6,\qtopheight){\circle*{2}}
  \put(0,\qtopheight){\line(1,0){12}}
\ifcase #1
    \put(6,0){\line(0,1){\qheight}}
\or \put(6,\qtopheight){\line(0,1){\qbottomheight}}
\or \put(6,\qtopheight){\line(0,-1){\qtopheight}}
\or {}
\or \put(1,0){\line(0,1){\qheight}}
    \qthicklines
    \put(11,0){\line(0,1){\qheight}}
    \thinlines
\or \qthicklines
    \put(1,0){\line(0,1){\qheight}}
    \thinlines
    \put(11,0){\line(0,1){\qheight}}
\fi
\end{picture}
}

\def\ex#1{
\begin{picture}(12,\qheight)(0,0)
  \put(6,\qtopheight){\circle{3}}
  \put(0,\qtopheight){\line(1,0){4.5}}
  \put(12,\qtopheight){\line(-1,0){4.5}}
\ifcase #1
    \put(6,0){\line(0,1){\qtopheightminusthreehalves}}
    \put(6,\qheight){\line(0,-1){\qbottomheightminusthreehalves}}
\or \put(6,\qheight){\line(0,-1){\qbottomheightminusthreehalves}}
\or \put(6,0){\line(0,1){\qtopheightminusthreehalves}}
\or {}
\or \put(1,0){\line(0,1){\qheight}}
    \qthicklines
    \put(11,0){\line(0,1){\qheight}}
    \thinlines
\or \qthicklines
    \put(1,0){\line(0,1){\qheight}}
    \thinlines
    \put(11,0){\line(0,1){\qheight}}
\fi
\end{picture}
}

\def\nt#1{
\begin{picture}(12,\qheight)(0,0)
  \put(6,\qtopheight){\circle{4}}
  \put(0,\qtopheight){\line(1,0){12}}
\ifcase #1
    \put(6,0){\line(0,1){\qheight}}
\or \put(6,\qheight){\line(0,-1){\qbottomheightplustwo}}
\or \put(6,0){\line(0,1){\qtopheightplustwo}}
\or \put(6,\qtopheightplustwo){\line(0,-1){4}}
\or \put(1,0){\line(0,1){\qheight}}
    \qthicklines
    \put(11,0){\line(0,1){\qheight}}
    \thinlines
    \put(6,\qtopheightplustwo){\line(0,-1){4}}
\or \qthicklines
    \put(1,0){\line(0,1){\qheight}}
    \thinlines
    \put(11,0){\line(0,1){\qheight}}
    \put(6,\qtopheightplustwo){\line(0,-1){4}}
\fi
\end{picture}
}

\def\ox#1{
\begin{picture}(12,\qheight)(0,0)
  \put(6,\qtopheight){\circle{3}}
  \put(0,\qtopheight){\line(1,0){12}}
\ifcase #1 
    \put(6,0){\line(0,1){\qheight}}
\or \put(6,\qheight){\line(0,-1){6.5}}
\or \put(6,0){\line(0,1){6.5}}
\fi
\end{picture}
}

\newcommand{\ct}[1]{
\begin{picture}(12,\qheight)(0,0)
  \multiput(0,\qtopheight)(11,0){2}{\line(1,0){1}}
  \put(6,\qtopheight){\circle{\qheight}}
  \put(0,0){\vbox to \myvsize{\vfill
	\hbox to \myhsize{\hfill #1\hfill}\vfill}}
\end{picture}
}

\newcommand{\ti}[1]{
\begin{picture}(12,\qheight)(0,0)
  \multiput(1,0)(\qheight,0){2}{\line(0,1){\qheight}}
  \multiput(1,0)(0,\qheight){2}{\line(1,0){\qheight}}
  \multiput(0,\qtopheight)(11,0){2}{\line(1,0){1}}
  \put(0,0){\vbox to \myvsize{\vfill
	\hbox to \myhsize{\hfill #1\hfill}\vfill}}
\end{picture}
}

\newcommand{\tc}[1]{
\begin{picture}(12,\qheight)(0,0)
 \put(6,\qtopheight){\circle{\qheight}}
 \multiput(0,\qtopheight)(11,0){2}{\line(1,0){1}}
 \put(0,0){\vbox to \myvsize{\vfill
	\hbox to \myhsize{\hfill #1\hfill}\vfill}}
\end{picture}
}

\newcommand{\tb}[2]{
\begin{picture}(12,\qheight)(0,0)
  \put(1,0){\line(0,1){\qheight}}
  \qthicklines
  \put(11,0){\line(0,1){\qheight}}
  \thinlines
  \multiput(1,0)(\qheight,0){2}{\line(0,1){\qheight}}
  \multiput(0,\qtopheight)(11,0){2}{\line(1,0){1}}
  \put(0,0){\vbox to \myvsize{\vfill
	\hbox to \myhsize{\hfill #2\hfill}\vfill}}
\ifcase #1
{}
\or \put(1,0){\line(1,0){\qheight}}
\or \put(1,\qheight){\line(1,0){\qheight}}
\or \multiput(1,0)(0,\qheight){2}{\line(1,0){\qheight}}
\fi
\end{picture}
}

\newcommand{\tp}[2]{
\begin{picture}(12,\qheight)(0,0)
  \qthicklines
  \put(1,0){\line(0,1){\qheight}}
  \thinlines
  \put(11,0){\line(0,1){\qheight}}
  \multiput(0,\qtopheight)(11,0){2}{\line(1,0){1}}
  \put(0,0){\vbox to \myvsize{\vfill
	\hbox to \myhsize{\hfill #2\hfill}\vfill}}
\ifcase #1
{}
\or \put(1,0){\line(1,0){\qheight}}
\or \put(1,\qheight){\line(1,0){\qheight}}
\or \multiput(1,0)(0,\qheight){2}{\line(1,0){\qheight}}
\fi
\end{picture}
}

\newcommand{\place}[1]{\vbox to \myvsize{\vfill
	\hbox to \myhsize{\hfill #1\hfill}\vfill}}
\def\plac#1#2{\vbox to \myvsize{\vfill
	\hbox to #1\myhsize{#2\hfill}\vfill}}

\definecolor{brown}{rgb}{0.6,0.4,0.2}
\definecolor{purple}{rgb}{0.8,0.0,1.0}
\definecolor{gray}{rgb}{0.5,0.5,0.5}

\title{Shor's Algorithm on a Nearest-Neighbor Machine}
\author{Samuel A. Kutin\thanks{Center for Communications Research, 805 Bunn Drive,
Princeton, NJ 08540. {\tt kutin@idaccr.org}}}
\date{}               

\newcommand{\caps}[1]{{\sc #1}}

\newcommand{\floor}[1]{\left\lfloor #1 \right\rfloor}
\newcommand{\ceil}[1]{\left\lceil #1 \right\rceil}
\newcommand{\xor}{\mathbin{\oplus}}
\newcommand{\xoreq}{\mathbin{\oplus\!=}}

\newcommand{\qu}[1]{{\left| {#1} \right\rangle}}
\newcommand{\phihat}{\smash[t]{\hat{\phi}}}

\newcommand{\DKRS}{cla}
\newcommand{\CDKM}{ripple}

\newcommand{\QFT}{\caps{QFT}\xspace}
\begin{document}

\maketitle
\begin{abstract}
We give a new ``nested adds'' circuit for implementing Shor's
algorithm in linear width and quadratic depth on a nearest-neighbor
machine.  Our circuit combines Draper's transform adder with
approximation ideas of Zalka.  The transform adder requires small
controlled rotations.  We also give another version, with slightly
larger depth, using only reversible classical gates.  We do not know
which version will ultimately be cheaper to implement.
\end{abstract}

\section{Introduction}
\label{intro-sec}

We describe a new quantum exponentiation circuit that obeys a
``nearest-neighbor'' constraint:  we imagine that qubits are arranged
in a line, and we are only allowed to perform interactions between
adjacent qubits.  Previous $n$-bit nearest-neighbor exponentiation
circuits~\cite{FDH,V}
required either depth $O(n^3)$ or superlinear width, but our construction
has width $O(n)$ and depth $O(n^2)$.  This new exponentiation circuit,
together with a nearest-neighbor quantum Fourier transform (QFT)~\cite{FDH},
gives a new circuit
for Shor's factorization algorithm~\cite{Shor}.

A number of people have constructed exponentiation circuits for general
architectures (i.e., without the nearest-neighbor restriction).
See, for example,~\cite{VMI,VMIL,V} for recent summaries.
Many of the techniques used to reduce circuit depth
do not appear to apply to a nearest-neighbor architecture.

Beauregard~\cite{Beau} has given a simple exponentiation
circuit using Draper's transform adder~\cite{Drap}.  The adder requires
two QFTs together with some controlled rotations.  Beauregard's circuit
uses only $2n + O(1)$ qubits, but has cubic depth---the dominant cost is
$\Theta(n^2)$ applications of the transform adder.
Fowler, Devitt, and Hollenberg~\cite{FDH} modify Beauregard's circuit for use on a
nearest-neighbor machine, and they show that these modifications do not
affect the dominant terms in the expression for size or depth.

Our contribution is a new approximate controlled modular multiplier with
linear width and linear depth.  We use
an idea of Zalka~\cite{Zalka} for building approximate multipliers.
While we still multiply by performing $O(n)$ additions, we only 
perform a constant number of large QFTs for each multiply.
When we insert our multiplier into the framework of Fowler et al.,
we obtain a nearest-neighbor exponentiation circuit with linear
width and quadratic depth.\footnote{Zalka~\cite{Z2} has recently pointed
out this same idea of performing mulitple additions framed by a single
QFT, but he does not work out any details or discuss the application
to nearest-neighbor circuits.}

We first set some notation and review prior work in Section~\ref{prelim-sec}.
We describe our multiplier and the resulting exponentiator
in Section~\ref{main-sec}, and we discuss a version for general
architectures in Section~\ref{general-sec}.

Following Fowler et al., we assume that any interaction between two
adjacent qubits has unit cost.  In practice, some gates may be easier
to implement than others.  Our circuit requires small controlled rotations
that may prove expensive.  Van Meter~\cite{V} discusses the error correction
requirements for various adders and suggests that the transform adder may not
be useful in practice.
In Section~\ref{classical-sec}
we describe a version of the circuit that is essentially classical
and that does not require these small rotations.  However, the
depth increases to $O(n^2 \log n)$.  This is the same asymptotic
depth achieved by Van Meter~\cite{V}, but we require only linear width.

\section{Preliminaries}
\label{prelim-sec}

Our goal is to compute $w = g^e \bmod m$.  Here $g$ and $m$
are $n$-bit constants, known to the classical compiler that builds
our circuit.  The $2n$-bit exponent $e$ is in quantum
memory.\footnote{More generally, $e$ has length $\alpha n$, and the
error rate of the algorithm depends on $\alpha$.  For simplicity
we take $\alpha = 2$.}  Using
a standard trick (see, for example,~\cite{Beau}),
we can assume that only one bit of
$e$ at a time is stored in our quantum computer.

Writing $e = \sum 2^i e_i$, we have 
$$
w = \left(\prod_i (g^{2^i} \bmod m)^{e_i}\right) \bmod m.
$$
That is, we can decompose our exponentiation into $2n$ controlled
multiplications.  In each case we multiply by $1$ if the controlling
bit $e_i$ is $0$, and by a constant if $e_i$ is 1.

In Section~\ref{prelim-mod-mult-sec}, we describe how we reduce
controlled modular multiplication to (roughly) $n$ controlled
additions.  In Section~\ref{prelim-transform-adder-sec}, we describe
the addition routine we will use.

We refer the reader to Fowler et al.~\cite{FDH} for
useful building blocks for nearest-neighbor circuits.  We will use
their ``mesh'' circuit for interleaving two registers.  We will
not use their controlled swap; instead, in Section~\ref{prelim-pseudo-sec}
we describe a simpler controlled swap for the case when one register is
known to be $0$.

\subsection{Approximate Modular Multiplication}
\label{prelim-mod-mult-sec}

We now present a scheme of Zalka~\cite{Zalka} for performing
controlled modular multiplication.  We wish to compute
$$
r = abc \bmod m,
$$
where $a$ and $m$ are $n$-bit constants, $b = \sum_i 2^i b_i$ is
in $n$ bits of quantum memory, and $c$ is a control bit.
We can write
$$
r \equiv abc \equiv \sum_i 2^i a b_i c \equiv \sum_i (b_i c) \left(2^i a \bmod m\right) \pmod m.
$$
We can view this as repeated controlled modular addition; the
numbers $x_i = 2^i a \bmod m$ are known at compile-time, and
we have $n$ control bits $y_i = b_i c$.

We define the partial sum
$$
s = \sum_i y_i x_i = r - qm.
$$
The sum $s$ is congruent to the answer $r \pmod m$.  Also, since 
$s < nm$, the quotient $q$ is at most $n$.  In particular, we can
write down $q$ using only $\log_2 n$ bits.

Zalka's key idea is to approximate the desired answer $r$ in two
parallel steps.  First, we compute $s$ by repeated controlled addition into
an $n$-bit accumulator.  Second,
we approximate $q$:   We choose some $\ell_0 = O(\log n)$, and we
compute $\hat{q}$ using only the $\ell_0$ high
bits of each $x_i$.  More precisely, let $\hat{x}_i = 2^{n-\ell_0}
\floor{x_i/2^{n-\ell_0}}$.  Then $\hat{q} = \floor{(\sum y_i \hat{x}_i) / m}$.
We can easily compute $\hat{q}$ in depth $O(\log^2 n)$.  With
high probability, $\hat{q} = q$.

Once we have $\hat{q} = \sum_i 2^i \hat{q}_i$, subtracting $\hat{q}m$
from $s$ can be done with $\log_2 n$ additional controlled adds into
our accumulator (we subtract $2^i m$ controlled by $\hat{q}_i$).
Next, we must erase $\hat{q}$; again; this takes only $O(\log^2 n)$
depth.  So, aside from a lower-order term, the cost of controlled
modular multiplication is about $n$ controlled additions,
or, equivalently, one controlled integer multiplication.

There are other schemes that give modular multiplication circuits
at a cost of three times the cost of integer multiplication (see,
for example,~\cite{Dhem}).  So it might seem that Zalka's idea would
save only a constant factor.  However, Zalka's idea is conceptually
simpler; without it, we might not have found the linear-depth
multiplier of Section~\ref{main-sec}.
 
\subsection{The Transform Adder}
\label{prelim-transform-adder-sec}

Most quantum arithmetic circuits are essentially classical in nature.
Draper~\cite{Drap} has given an addition circuit that is inherently
quantum.  We briefly describe this circuit, and then discuss how to
adapt it to the nearest-neighbor setting.

Suppose we have an $n$-bit number register containing
$u = \sum_{j=0}^{n-1} u_j 2^j$.  Then the {\QFT} maps $\qu{u}$ to
$$
\qu{\phi(u)} =
\frac{1}{2^{n/2}}\sum_{k=0}^{2^n - 1} e^{2 \pi i u k / 2^n} \qu{k}
= \bigotimes_{j=0}^{n-1} \qu{\phi_j(u)},
$$
where
$$
\phi_j(u) = {1 \over \sqrt{2}} \left(\qu{0} + e^{2 \pi i u / 2^{j+1}}\qu{1}
\right).
$$
Note that $\qu{\phi(u)}$ is an unentangled state.

Suppose we want to add $v$ to $u$.  We can 
replace each bit $\phi_j(u)$ by $\phi_j(u + v)$; this is simply a
$Z$-rotation by an angle of $2 \pi v / 2^{j+1}$, so we can rotate each
bit independently.  To perform controlled addition, each of these
rotations is controlled by a bit $c$.  We can then perform an inverse
{\QFT} to change $\qu{\phi(u+v)}$ to $\qu{u+v}$.

One way to view the {\QFT} is that we have moved the information about
$u$ into the phase of the qubits.  To do a modular reduction and test
the high bit of $u$, we first need to perform an inverse {\QFT}.
So, for a naively designed modular exponentiation circuit, we perform
$\Theta(n^2)$ {\QFT}s and inverse {\QFT}s.
Our main result is
a circuit design with only $O(n)$ {\QFT}s.

\begin{figure}[h]
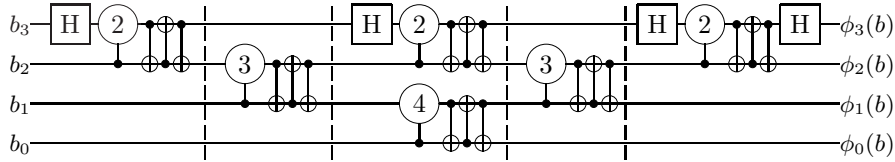

\begin{center}
\renewcommand{\arraystretch}{0}
\begin{tabular}{r@{}*{30}{c@{}}l}
\wire{$b_3$} &\sn0&\ti H&\tc{2}&\cn2&\xn2&\cn2&\sm3&\sx0&\sn0&\sn0&\sn0&\sm3&\ti H&\tc{2}&\cn2&\xn2&\cn2&\sm3&\sx0&\sn0&\sn0&\sn0&\sn3&\ti H&\tc{2}&\cn2&\xn2&\cn2&\ti H&\sn0& \wire{$\phi_3(b)$} \\
\wire{$b_2$} &\sn0&\sx0&\dx1&\xn1&\cn1&\xn1&\sm3&\tc{3}&\cn2&\xn2&\cn2&\sm3&\sx0&\dx1&\xn1&\cn1&\xn1&\sm3&\tc{3}&\cn2&\xn2&\cn2&\sn3&\sx0&\dx1&\xn1&\cn1&\xn1&\sx0&\sn0& \wire{$\phi_2(b)$} \\
\wire{$b_1$} &\sn0&\sx0&\sx0&\sn0&\sn0&\sn0&\sm3&\dx1&\xn1&\cn1&\xn1&\sm3&\sx0&\tc{4}&\cn2&\xn2&\cn2&\sm3&\dx1&\xn1&\cn1&\xn1&\sn3&\sx0&\sx0&\sn0&\sn0&\sn0&\sx0&\sn0& \wire{$\phi_1(b)$} \\
\wire{$b_0$} &\sn0&\sx0&\sx0&\sn0&\sn0&\sn0&\sm3&\sx0&\sn0&\sn0&\sn0&\sm3&\sx0&\dx1&\xn1&\cn1&\xn1&\sm3&\sx0&\sn0&\sn0&\sn0&\sn3&\sx0&\sx0&\sn0&\sn0&\sn0&\sx0&\sn0& \wire{$\phi_0(b)$} \\
\end{tabular}

\end{center}
\caption{Quantum Fourier transform of a 4-bit register on a
nearest-neighbor machine. \textcircled{\scriptsize$j$}
denotes a $Z$-rotation by
$2 \pi / 2^j$.}
\label{qft-fig}
\end{figure}

Fowler et al.~\cite{FDH} give a nearest-neighbor
form of the {\QFT}.  A 4-bit version is depicted
in Figure~\ref{qft-fig}.  After each controlled rotation, we swap the
two bits involved, so every pair of bits can interact.  (If we leave out
the swaps, we obtain the linear-depth {\QFT} of Moore and Nilsson~\cite{MN}.)
Note that we
assign unit cost to the controlled rotation together with the
accompanying swap.

The size of this {\QFT} circuit is $n^2/2 + O(n)$.  We may be able to
approximate the {\QFT} and skip some of the small rotations.  On a
general machine, this reduces the size to $O(n \log n)$, but on a
nearest-neighbor machine we still have to perform $n \choose 2$
swaps.

\subsection{Pseudo-Toffolis and Controlled Swaps}
\label{prelim-pseudo-sec}

\begin{figure}[h]
\begin{center}
\renewcommand{\arraystretch}{0}
\begin{tabular}{r@{}*{11}{c@{}}l}
\wire{$u$} &\sn0&\cn2&\sm0& \wire{${}$} &\sn0&\dx2&\sn0&\sx0&\sn0&\dx2&\sn0& \\
\wire{$v$} &\sn0&\xn0&\sm0& \wire{$\quad\cong\quad$} &\sn0&\ti H&\sn0&\ti Z&\sn0&\ti H&\sn0& \\
\wire{$w$} &\sn0&\cn1&\sm0& \wire{${}$} &\sn0&\sx0&\sn0&\dx1&\sn0&\sx0&\sn0& \\
\end{tabular}

\end{center}
\caption{Pseudo-Toffoli gate $v \xoreq uw$.  We also change
the phase when $\qu{uvw} = \qu{011}$.}
\label{pseudo-fig}
\end{figure}

A frequent useful building block for our circuit is a {\em Toffoli\/}
gate, or doubly-controlled not: $v \xoreq uw$.  A cascade of
Toffoli gates through a $k$-bit register has depth $2k$.  However,
if we use the ``pseudo-Toffoli'' gate of Figure~\ref{pseudo-fig},
the depth of the cascade can be reduced to $k$.
See~\cite{BBCDMSSSW} for an equivalent pseudo-Toffoli gate.

The idea of Figure~\ref{pseudo-fig} is that we correctly set $v$ to
$v \xor uw$, but we change the phase when $\qu{uvw} = \qu{011}$.
Normally this would be an unacceptable side effect, but there are
two cases where we are okay:  First, we may plan to undo this
computation and fix the phase later.  Second, we may know that the
problem input is forbidden for some reason.

\begin{figure}
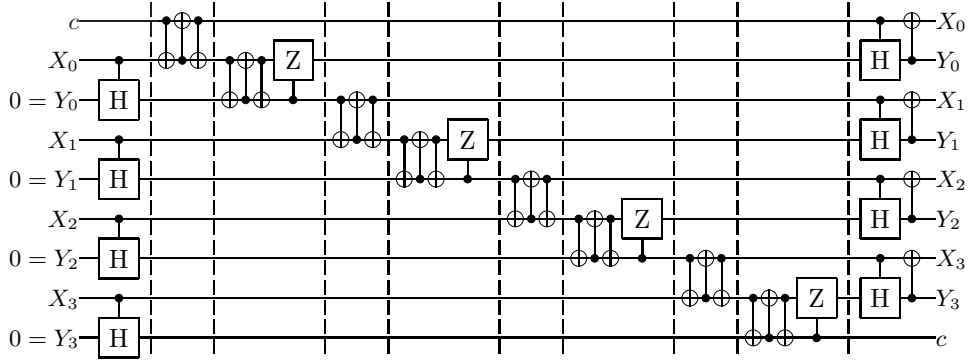

\begin{center}
\renewcommand{\arraystretch}{0}
\begin{tabular}{r@{}*{42}{c@{}}l}
\wire{$c$} &\sn0&\sx0&\sn3&\cn2&\xn2&\cn2&\sn3&\sn0&\sn0&\sn0&\sx0&\sn3&\sn0&\sn0&\sn0&\sn3&\sn0&\sn0&\sn0&\sx0&\sn3&\sn0&\sn0&\sn0&\sn3&\sn0&\sn0&\sn0&\sx0&\sn3&\sn0&\sn0&\sn0&\sn3&\sn0&\sn0&\sn0&\sx0&\sn3&\dx2&\xn2&\sn0& \wire{$X_0$} \\
\wire{$X_0$} &\sn0&\dx2&\sn3&\xn1&\cn1&\xn1&\sn3&\cn2&\xn2&\cn2&\ti Z&\sn3&\sn0&\sn0&\sn0&\sn3&\sn0&\sn0&\sn0&\sx0&\sn3&\sn0&\sn0&\sn0&\sn3&\sn0&\sn0&\sn0&\sx0&\sn3&\sn0&\sn0&\sn0&\sn3&\sn0&\sn0&\sn0&\sx0&\sn3&\ti H&\cn1&\sn0& \wire{$Y_0$} \\
\wire{$0=Y_0$} &\sn0&\ti H&\sn3&\sn0&\sn0&\sn0&\sn3&\xn1&\cn1&\xn1&\dx1&\sn3&\cn2&\xn2&\cn2&\sn3&\sn0&\sn0&\sn0&\sx0&\sn3&\sn0&\sn0&\sn0&\sn3&\sn0&\sn0&\sn0&\sx0&\sn3&\sn0&\sn0&\sn0&\sn3&\sn0&\sn0&\sn0&\sx0&\sn3&\dx2&\xn2&\sn0& \wire{$X_1$} \\
\wire{$X_1$} &\sn0&\dx2&\sn3&\sn0&\sn0&\sn0&\sn3&\sn0&\sn0&\sn0&\sx0&\sn3&\xn1&\cn1&\xn1&\sn3&\cn2&\xn2&\cn2&\ti Z&\sn3&\sn0&\sn0&\sn0&\sn3&\sn0&\sn0&\sn0&\sx0&\sn3&\sn0&\sn0&\sn0&\sn3&\sn0&\sn0&\sn0&\sx0&\sn3&\ti H&\cn1&\sn0& \wire{$Y_1$} \\
\wire{$0=Y_1$} &\sn0&\ti H&\sn3&\sn0&\sn0&\sn0&\sn3&\sn0&\sn0&\sn0&\sx0&\sn3&\sn0&\sn0&\sn0&\sn3&\xn1&\cn1&\xn1&\dx1&\sn3&\cn2&\xn2&\cn2&\sn3&\sn0&\sn0&\sn0&\sx0&\sn3&\sn0&\sn0&\sn0&\sn3&\sn0&\sn0&\sn0&\sx0&\sn3&\dx2&\xn2&\sn0& \wire{$X_2$} \\
\wire{$X_2$} &\sn0&\dx2&\sn3&\sn0&\sn0&\sn0&\sn3&\sn0&\sn0&\sn0&\sx0&\sn3&\sn0&\sn0&\sn0&\sn3&\sn0&\sn0&\sn0&\sx0&\sn3&\xn1&\cn1&\xn1&\sn3&\cn2&\xn2&\cn2&\ti Z&\sn3&\sn0&\sn0&\sn0&\sn3&\sn0&\sn0&\sn0&\sx0&\sn3&\ti H&\cn1&\sn0& \wire{$Y_2$} \\
\wire{$0=Y_2$} &\sn0&\ti H&\sn3&\sn0&\sn0&\sn0&\sn3&\sn0&\sn0&\sn0&\sx0&\sn3&\sn0&\sn0&\sn0&\sn3&\sn0&\sn0&\sn0&\sx0&\sn3&\sn0&\sn0&\sn0&\sn3&\xn1&\cn1&\xn1&\dx1&\sn3&\cn2&\xn2&\cn2&\sn3&\sn0&\sn0&\sn0&\sx0&\sn3&\dx2&\xn2&\sn0& \wire{$X_3$} \\
\wire{$X_3$} &\sn0&\dx2&\sn3&\sn0&\sn0&\sn0&\sn3&\sn0&\sn0&\sn0&\sx0&\sn3&\sn0&\sn0&\sn0&\sn3&\sn0&\sn0&\sn0&\sx0&\sn3&\sn0&\sn0&\sn0&\sn3&\sn0&\sn0&\sn0&\sx0&\sn3&\xn1&\cn1&\xn1&\sn3&\cn2&\xn2&\cn2&\ti Z&\sn3&\ti H&\cn1&\sn0& \wire{$Y_3$} \\
\wire{$0=Y_3$} &\sn0&\ti H&\sn3&\sn0&\sn0&\sn0&\sn3&\sn0&\sn0&\sn0&\sx0&\sn3&\sn0&\sn0&\sn0&\sn3&\sn0&\sn0&\sn0&\sx0&\sn3&\sn0&\sn0&\sn0&\sn3&\sn0&\sn0&\sn0&\sx0&\sn3&\sn0&\sn0&\sn0&\sn3&\xn1&\cn1&\xn1&\dx1&\sn3&\sx0&\sn0&\sn0& \wire{$c$} \\
\end{tabular}

\end{center}
\caption{Swap of 4-bit registers $X$ and $Y$ controlled by $c$
in depth $10$.  We assume that $Y$ is initialized to $0$.}
\label{pseudo-cascade-fig}
\end{figure}

For example, suppose we want to swap two $n$-bit registers
$X$ and $Y$ controlled by a bit $c$.  Suppose further that $Y$ is
initialized to $0$.  Then we can build a pseudo-Toffoli cascade
as in Figure~\ref{pseudo-cascade-fig}.  Since each Toffoli target is
known to be $0$, there will be no phase shift.  The depth is $2n + 2$.

\section{Nested Adds}
\label{main-sec}

We now describe our main result, the ``nested adds'' multiplier.
We begin by describing a controlled multiplier with linear width
and depth; we then explain how to modify it to be a modular multiplier.
We conclude with an exponentiation circuit with linear width and
quadratic depth.

\subsection{Nested Controlled Addition}
\label{main-add-sec}

As noted in Section~\ref{prelim-mod-mult-sec}, we can view
controlled multiplication as repeated controlled addition.
In this section, we build a repeated controlled adder.
We have an $n$-bit
register $Z$, initialized to some value $z$, and an $n$-bit
register $Y$ of control bits $y_i$.  When the circuit concludes,
we want $Z$ to contain $$\left(z + \sum_i x_i y_i\right) \bmod 2^n,$$ where
the values $x_i$ are $n$-bit constants.  In the next section, we
will convert this circuit to a modular multiplier.

It is clear that $n$-bit addition controlled by a single bit $y_i$
requires linear depth on a nearest-neighbor machine; the control
bit can affect all $n$ bits of $Z$, so we need linear time to
move (or pseudocopy) it from one end to the other.  One might at
first think that performing $n$ controlled additions would require
quadratic depth.  However, if we use the transform adder, we can
nest the additions.

\begin{figure}[h!]
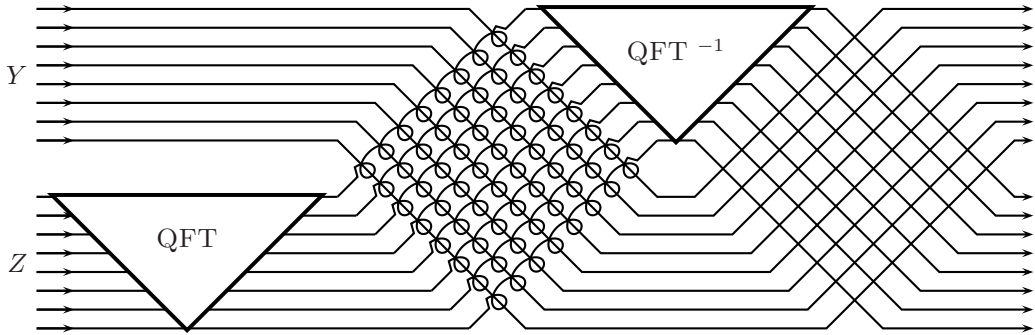

\begin{center}
\input nested.pst
\end{center}
\caption{Schematic for the ``nested adds'' repeated controlled adder.}
\label{nested-fig}
\end{figure}

The basic structure of the circuit is depicted in Figure~\ref{nested-fig}.
We begin by performing the {\QFT} on $Z$, in depth $2n-3$.
Next, we take each bit of $Y$ successively and swap it with each
bit of $Z$.  As we swap $Y_i$ with $Z_j$, we also rotate $Z_j$
controlled by $Y_i$; the rotation amount depends on $x_i$.  The idea
is that we are adding in $x_i$ by rotating each bit of $Z$ by the
proper amount; all of these rotations commute, so the order is
unimportant.  This portion has depth $2n - 1$; when it concludes,
we have effectively swapped the $Z$ and $Y$ registers.

Next, we perform the inverse \QFT on $Z$.  This again has depth $2n-3$.
Finally, we move $Y$ back to where it started in depth $2n - 1$.

As described, the total depth would be $8n - 8$.  However, as shown
in Figure~\ref{nested-fig}, the inverse \QFT nests nicely with the
swaps with $Y$.  We can start the inverse \QFT at time $3n - 5$, and
we can start the final swaps at time $4n-2$.  The total depth is only
$6n - 4$.

If we can assume $z$ is a constant, then we can replace the initial {\QFT}
with a single time-slice of $n$ unitary transformations\footnote{For
example, when $z = 0$, we apply a Hadamard to each qubit of $Z$.} on $Z$.
The depth is reduced to $4n - 1$.  See Section~\ref{main-error-sec} for
the reasons
why we might want to allow nonzero $z$.  For the remainder of this paper,
we will assume that $z$ is a constant, and that we can skip the initial {\QFT}.

\subsection{Nested Controlled Modular Addition}
\label{main-mod-mult-sec}

To turn the above circuit into a modular multiplier, we follow the
procedure described in Section~\ref{prelim-mod-mult-sec}.  We
compute the sum $s = \sum_i y_i x_i$ congruent to the desired
answer $r$ modulo $m$.  (Since we know our final answer has $n$ bits, we
need only compute the low $n$ bits of $s$.)
Simultaneously, we compute the approximate
quotient $\hat{q}$.  We then subtract $\hat{q}m$ from our main register.
Finally, we erase $\hat{q}$.

We compute $\hat{q}$ in an $\ell$-bit register $Q$, which we
locate between $Y$ and $Z$.  We take $\ell = \ell_0 + \log_2 n$, so
we have room to write the $(n + \log_2 n)$-bit sum $\sum_i y_i \hat{x}_i$
(which has $0$ in the low-order $n - \ell_0$ bits).

We need to initialize the low $\ell_0$ bits of $Q$.  If we have
nonconstant data in $Z$, we could pseudocopy
$\ell_0$ bits of it to $Q$; this is
not expensive, but it might be costly to erase $Z$ when we are done.
In our case, we will initialize $Z$ to a constant $z$, and $Q$
to the high-order $\ell_0$ bits of $z$.

We pass the bits of $Y$ past $Q$ and
then $Z$.  We compute the high bits of $z + \sum_i y_i \hat{x}_i$ in $Q$,
and we compute $z + \sum y_i x_i \bmod 2^n$ in $Z$.  

As soon as the last $y_i$ bit has passed through $Q$, we compute $\hat{q}$.
For $k = \log_2 n$ down to $1$, we first subtract $2^{k-1} m$ from
$Q$ by doing a unary rotation on each bit.  Next, we do an inverse \QFT in
depth at most $2\ell-1$;
the top bit of $Q$ is now a control bit indicating whether
we should have subtracted $2^{k-1} m$ or not.  We label that bit $\hat{q}_k$
and think of it as no longer part of $Q$.  We now do a \QFT on the
remaining bits of $Q$, and then move $\hat{q}_k$ through $Q$; this adds
$2^{k-1}m$ back if necessary, and also positions $\hat{q}_k$ to go through
$Q$.

At step $k$, we perform an inverse \QFT on $\ell_0 + k$ bits and
a \QFT on $\ell_0 + k - 1$ bits, and then we move $\hat{q}_k$ through $Q$.
The depth is $4(\ell_0 + k) - 3$.  The total depth, summing from
$k = 1$ to $\log_2 n$, is
\begin{equation}
\label{q-time}
2\ell^2 - 2\ell_0^2 + O(1) = 2 (2\ell - \log_2 n) \log_2 n + O(1).
\end{equation}

We use the $\hat{q}_k$ bits as control bits, subtracting $2^k m$ as
needed from $s$.  When we are done, the answer $r$ is in $Z$.  When we
pass the $\hat{q}_k$ bits back up, we again take time given
by~\eqref{q-time} to uncompute $\hat{q}$.  (Alternatively, we could
move all of $Q$ past $Z$ and then uncompute $\hat{q}$.)

We subtract $z$ from $Z$ after computing $r$.  See
Section~\ref{main-error-sec} for details.

The total circuit depth for repeated controlled addition is
$$
4n + 4 (2\ell - \log_2 n) \log_2 n + O(\log n).
$$
The width is $2n + \ell + O(1)$.

\subsection{Controlled Modular Multiplication}
\label{main-control-sec}

So far, we have assumed that the $n$ control bits are present at the
start of the computation.  To complete our modular multiplier, we need
to explain how to start from the multiplicand $b$ and
the overall control bit $c$ and produce the control bits $y_i = b_i c$.
Also, since we want an in-place multiplier, we need to explain how to
erase $b$ when we are done (if $c=1$).

\begin{sidewaysfigure}
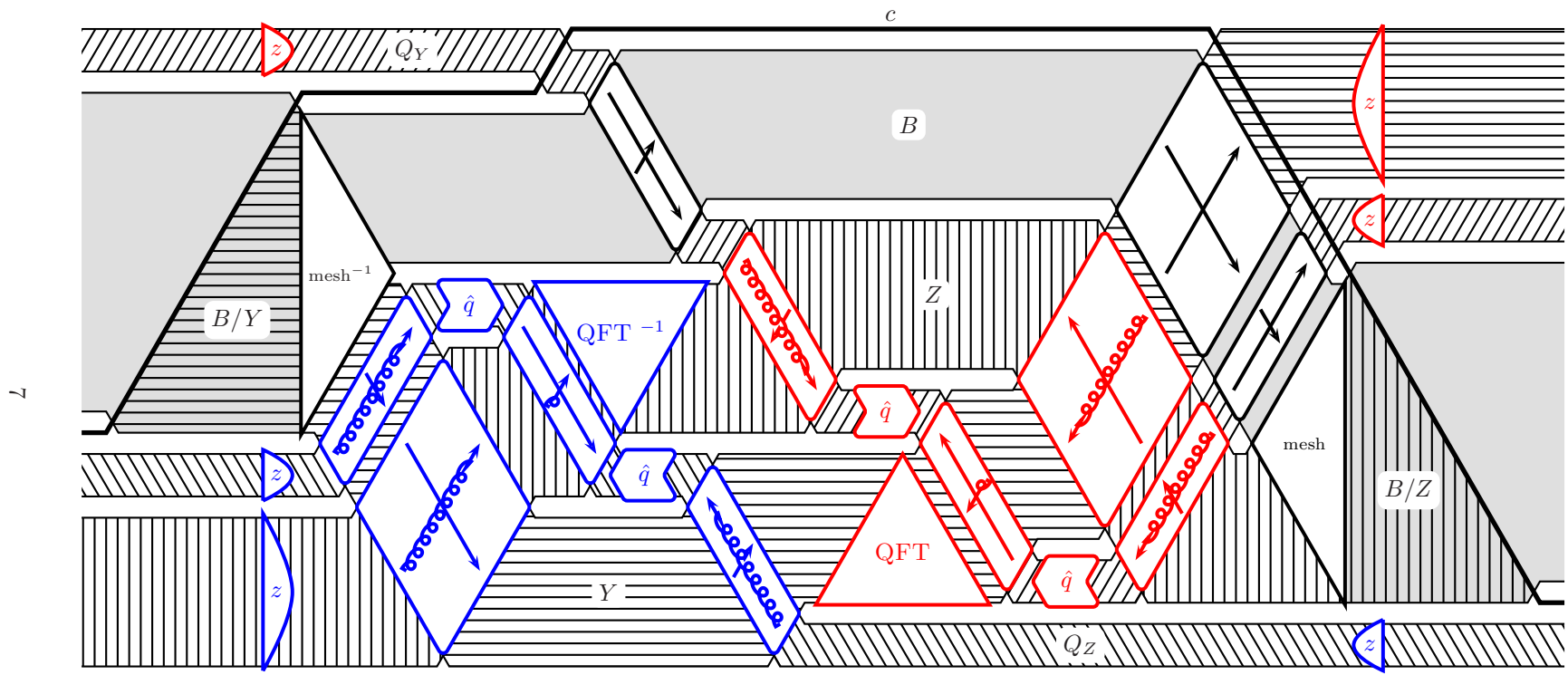

\begin{center}
\input mult.pst
\end{center}
\caption{Schematic for the ``nested adds'' controlled in-place
modular multiplier.}
\label{nested-mod-mult-fig}
\end{sidewaysfigure}

It is easy to perform the desired steps in linear depth, given
the linear-depth out-of-place modular multiplication
circuit described above.  The challenging part is to keep the
depth as low as possible.  Our solution has depth
$$
11n + 6 (2\ell - \log_2 n) \log_2 n + O(\log n),
$$
width
$$
3n + 2\ell + 1,
$$
and size
$$
5n^2 + O(n \log n),
$$
and is depicted in Figure~\ref{nested-mod-mult-fig}.  We briefly
describe the basic features of the circuit.

We have three $n$-bit registers (labeled $B$, $Y$, and $Z$),
two $\ell$-bit registers (labeled $Q_Y$ and $Q_Z$), and one
control bit $c$.  Initially $B$ contains $b$ and the other
four registers contain $0$.  When the circuit concludes,
$B$ contains $b$ (when $c=0$) or $ab$ (when $c=1$) and the
other four registers contain $0$.

To start, we have $Q_Y$, then $B$ and $Y$ interleaved (i.e.,
we have $B_0$, $Y_0$, $B_1$, $Y_1$, \dots, $B_{n-1}$, $Y_{n-1}$),
and then $c$, $Q_Z$, and $Z$.  When the circuit completes,
we have $Y$, then $Q_Y$, then $B$ interleaved with $Z$, then
$c$, and finally $Q_Z$.  So, except for the location of $c$,
the bits have been flipped upside-down.  (See
Section~\ref{main-exp-sec} for the reason we end with $c$ in a
different place.)

We first move $c$ through the interleaved $B$ and $Y$,
performing controlled swaps.  If the contents of $B$ and $Y$ were
wholly general, this process would have depth $4n$, but because
we know $Y$ contains $0$ we can use pseudo-Toffolis (see
Section~\ref{prelim-pseudo-sec}), and the depth is only $2n+2$.
After the controlled swaps, we unmesh $B$ and $Y$.

Next, we multiply $Y$ by $a$ and write the result to $Z$.
These gates are depicted in blue in Figure~\ref{nested-mod-mult-fig}.
We use $Q_Z$ as a scratch register for computing $\hat{q}$.  We
load a constant $z$ into $Z$ (and its high bits into $Q_Z$), then
we perform the circuit described in the previous section, and
finally we erase $Q_Z$ and unload the constant $z$.  When this
portion concludes, if $c = 0$, then $B$ contains $b$ and $Y$ and
$Z$ contain $0$.  If $c = 1$, then $B$ contains $0$, $Y$ contains
$b$, and $Z$ contains $ab$.

We now perform the gates depicted in red in
Figure~\ref{nested-mod-mult-fig}.  We undo a multiplication
of $Z$ by $a^{-1}$, writing the result into $Y$.  The red circuit
is a backwards, upside-down version of the blue circuit.  When
we are done, $Y$ contains $0$.  If $c = 0$, then $B$
contains $b$ and $Z$ contains $0$; if $c = 1$, then $B$
contains $0$ and $Z$ contains $ab$.

Finally, we mesh $B$ and $Z$ and perform the controlled swap
in reverse.  (Again, we can use pseudo-Toffolis to reduce the
depth to $2n+2$.)  We write $b$ or $ab$ to $B$, and we write $0$
to $Z$, as desired.

Note that part of the red circuit overlaps part of the blue
circuit.  In particular, we uncompute the first $\hat{q}$
while computing the second.  This is why the second-order
term in the depth is $6 (2\ell -\log_2 n)\log_2 n$ rather than
$8 (2\ell -\log_2 n)\log_2 n$.

We must swap $B$ and $Y$ before we can interleave
$B$ and $Z$.  If our bits were arranged in a ring, we could
bring $B$ around from the other side; this would reduce
the depth by about $n$ and the size by about $n^2$.  One
could construct a more symmetric version of
Figure~\ref{nested-mod-mult-fig} by moving $B$ down to the bottom
between the blue and red portions, but this increases the
size by about $n^2$ without changing the depth.

\subsection{Exponentiation}
\label{main-exp-sec}

We recall from Section~\ref{prelim-sec} that our goal is to
perform $2n$ controlled in-place modular multiplications.  We
will repeatedly apply the circuit of Section~\ref{main-control-sec}.
Since that circuit leaves the machine ``upside-down,'' we alternate
between applying the circuit right-side-up and upside-down.

Let $e_i$ denote the control bit in the $i$th round.  We add one
additional bit to the circuit of Section~\ref{main-control-sec}.
Just before we start the swap of $B$ and $Z$ controlled by $e_i$, we
create our next control bit $e_{i+1}$.  Then, as soon as we have
swapped two bits of the interleaved $B$ and $Z$ controlled by
$e_i$, we swap them again controlled by $e_{i+1}$ (viewing them
as $B$ and $Y$ for the next round).  We can thus overlap these
two controlled swaps; we reduce the depth of each round to only
$9n + O(\log^2 n)$.

There may be a technicality here because of the order in which we
perform measurements.  After we are done using $e_i$, we measure
it, and we may need to rotate $e_{i+1}$ based on the observed
value of $e_i$.  We will assume that this is not a problem in
practice.  If necessary, we could generate $\Theta(\sqrt{n})$
control bits at a time and use them; we would still have a
depth of roughly $9n$ and a width of roughly $3n$.

Our circuit has depth
\begin{gather*}
18 n^2 + 12 n (2\ell -\log_2 n) \log_2 n + O(n \log n),
\intertext{width}
3n + 2\ell + 2,
\intertext{and size}
10 n^3 + O(n^2 \log n).
\end{gather*}
Here $\ell = O(\log n)$ is chosen to control
the error rate of our computation of $\hat{q}$.  See the
next section for details.

\subsection{Error Analysis}
\label{main-error-sec}

In this section we address two questions.  First, how should
we choose $\ell$?  Second, how does filling $Z$
with a random value $z$ improve our error analysis?

We perform $4n$ modular multiplications.  For each of these, we
add $n$ quantities to compute $\hat{q}$.  There are thus $4n^2$
additions where we might make a mistake.  Given random addends,
the probability of an error propagating across a window of length
$\ell_0$ is $2^{-\ell_0}$.  Our probability of making an error
is therefore at most
$$
4n^2 2^{-\ell_0} = 2^{2 \log_2 n + 2 - \ell_0}.
$$
To reduce our error probability to a constant, we should take
$\ell_0 = 2 \log_2 n + O(1)$, or
$$\ell = \ell_0 + \log_2 n = 3 \log_2 n + O(1).$$

What does an error rate of $\epsilon$ mean in the quantum setting?
Instead of attaining the desired state $\qu{\phi}$, we attain a
state $\qu{\phihat} = \alpha \qu{\phi} + \eta \qu{\psi}$, where
the error state $\qu{\psi}$ is orthogonal to
$\qu{\phi}$ and $|\eta|^2 \le \epsilon$.
A standard calculation yields that the distance between the probability
distributions on measurements for
$\qu{\phi}$ and $\qu{\phihat}$ is at most $\epsilon$.
Note that an error may mean that we fail to erase scratch space
correctly, invalidating future rounds, but this is irrelevant to
the analysis.

The assumption above of ``random addends'' may not be reasonable.
Zalka~\cite{Zalka} discusses this problem: citing a ``private
objection'' by Manny Knill, Zalka writes that ``mathematically (and
therefore very cautiously) inclined people have questioned the
validity of this assumption.''  Our solution is to fill our
register with a random constant $z$.  (We can use the same $z$ each
time, or we can choose a different one for each multiplication.)
The expected probability of an error in computing $\hat{q}$ over
all our choices of $z$ is the desired $\epsilon$.

However, the constant $z$ introduces another place where errors can
occur.  When we subtract $z$ at the end, we do not perform a modular
subtraction.  If we ensure $z < m/2^t$, the probability of an error
at some point is $4n 2^{-t}$.  We therefore take $t = \log_2 n +
O(1)$.  Note that this increases $\ell_0$ to $3 \log_2 n + O(1)$
and $\ell$ to $4 \log_2 n + O(1)$.

\section{A Classical Version}
\label{classical-sec}

The circuit of this paper requires numerous small controlled rotations.
We now show that a variant of these ideas gives a reversible classical
approximate exponentiation circuit with depth $O(n^2 \log n)$ and
size $O(n^3)$.

We still organize exponentiation as repeated multiplication and
multiplication as repeated addition.  On a general architecture, we
can attain depth $O(n^2 \log n)$ using a logarithmic-depth
adder~\cite{\DKRS}.  On a nearest-neighbor machine, we cannot
perform controlled addition in sublinear depth.  As in our main
construction, we nest different controlled additions to obtain an
amortized depth of $O(\log n)$ per addition.

We return to the setting of Section~\ref{main-add-sec}.  We
have an $n$-bit register $Z$ (initialized to some value $z$) and
an $n$-bit register $Y$.  We wish to write to $Z$ the quantity
$z + \sum_i x_i y_i \bmod 2^n$; here the $y_i$s are bits of $y$ and
the $x_i$s are $n$-bit constants.

We follow the general structure of Figure~\ref{nested-fig}.  Since
we wish to build a classical circuit, we no longer perform any
{\QFT}s.  Instead, we choose some $t = O(\log n)$, and we write
$k = \ceil{n/t}$.  We divide $Z$ into $k$ blocks of size $t$; each
``wire'' of $Z$ in Figure~\ref{nested-fig} represents a single block
$Z^j$.  (Each wire of $Y$ is still a single bit $y_i$.)  We also
divide each $x_i$ into blocks $X_i^j$ of length $t$.

We divide this portion of the circuit into $n+k-1$ rounds.  In
round $r$, $y_{r-j}$ crosses $Z_j$ for all $j$ (as long as $0 \le j < k$
and $0 \le r-j < n$).  At this time, we add the number
$$
A_r = \sum_j y_{r-j} X_{r-j}^j 2^{t(j-1)}
$$
into $Z$.  Note that
$$\sum_{r=0}^{n+k-1} A_r = \sum_{i=0}^{n-1} x_i y_i$$
as desired.  Also note that, in round $r$, the control bit
$y_{r-j}$ controlling the $j$th block of $A_r$ is next to $Z_j$ in
memory.

To add $A_r$ into $Z$, we first do $k$ parallel controlled adds, one
for each block.  We erase our work, but we write down the high bit
$h_j$ for each block.  We hope that we correctly compute each $h_j$;
this requires that no carry propagate through an entire block.

Next, we again do $k$ parallel controlled adds, but this time, for
the $j$th block, we use $h_{j-1}$ as an incoming carry bit.  If
the $h_j$ bits are all correct, we correctly add $A_r$ into $Z$.

Finally, we erase the $h_j$ bits.  We compare $Z_j$ with
$y_{r-j} X_{r-j}^j$ to determine if an overflow occurred; if so,
$h_j$ must have been $1$.  We then exchange each $y_{r-j}$ bit with
$Z_j$ to move the control bits into position for the next round.

Each of these steps can be performed with a ripple-carry
adder~\cite{\CDKM}; the depth is $Ct$ for a small constant $C$.  We need $2k$
extra bits:\ the high bits $h_j$ and one scratch bit for each
ripple.\footnote{We cannot use the ripple-carry adder of Takahashi
and Kunihiro~\cite{TK}.  Their adder eliminates the scratch bit,
but it does not work on a nearest-neighbor machine.}

To do modular multiplication, we use the same scheme as in our
main construction: we estimate $\hat{q}$ on the side.  The error
analysis is the same.  Note that we also perform $O(n^3)$
controlled additions of size $t$;
the probability that some $h_j$ bit is wrong at
some point is thus $O(n^3 2^{-t})$.  We choose $t = O(\log n)$ to
reduce this probability to a small constant.

We can use the pseudo-Toffoli
gates described in Section~\ref{prelim-pseudo-sec} to reduce the
depth.  It is interesting to note that, for the ripple-carry adder,
we do not perform exactly the same gates when we undo the computation,
but the ``bad'' case for the pseudo-Toffoli happens on the forward
ripple if and only if it happens on the reverse ripple, so we fix
our phase errors correctly.

The circuit depth is $O(n^2 \log n)$.  The exact constant depends
on the choice of $\ell$ and $t$ and on precisely how we
perform the ripple-carry additions.

\section{General Architectures}
\label{general-sec}

The ``nested adds'' multiplier of Section~\ref{main-sec} can be
simplified in several ways if implemented on a machine without
a nearest-neighbor restriction:
\begin{itemize}
\item The controlled swaps at the start and end of the multiplier can
be performed in logarithmic depth.  We fan the control bit $c$ out into
an empty $n$-bit register, perform $n$ parallel swaps, and fan $c$
back in.  Note that we always have an empty $n$-bit register available.
\item The mesh and unmesh operations and any register swaps (all in
black in Figure~\ref{nested-mod-mult-fig}) are unnecessary.  This
reduces the depth by about $n$ and the size by about $2n^2$.
\item The {\QFT} and inverse {\QFT} can be approximated.  This does
not improve the depth, but the size of each decreases from
about $n^2 / 2$ to $O(n \log n)$.
\end{itemize}

With these changes, the modular multiplier has depth $6n +
6 (2\ell -\log_2 n)\log_2 n
+ O(\log n)$, width $3n + 2\ell + 1$, and size $2n^2 + O(n \log n)$.
Taking $\ell = 3 \log_2 n + O(1)$ as in Section~\ref{main-error-sec},
we get an exponentiation circuit with depth
\begin{gather*}
12n^2 + 60 n \log_2^2 n + O(n \log n),
\intertext{width}
3n + 6 \log_2 n + O(1),
\intertext{and size}
4n^3 + O(n^2 \log n).
\end{gather*}

We could further reduce the depth by using a parallel version of
the {\QFT}~\cite{CW}, but each multiply would still have depth at least
$5n + O(\log^2 n)$.
We could also consolidate the registers $Q_Y$ and $Q_Z$; we would
get a slight increase in depth and a slight decrease in width.

\section*{Acknowledgements}
The author thanks Bob Beals, Tom Draper, and David Moulton for
numerous discussions.

\bibliography{nn}
\bibliographystyle{alpha}

\end{document}